# Cellular Automata Rules and Linear Numbers

Birendra Kumar Nayak, Sudhakar Sahoo and Sagarika Biswal

Institute of Mathematics and Applications,
Andharua, Bhubaneswar-751003
Email: bknatuu@yahoo.co.uk, sudhakar.sahoo@gmail.com and sagarika_bbs@yahoo.co.in

*Abstract:*- In this paper, linear Cellular Automta (CA) rules are recursively generated using a binary tree rooted at "0". Some mathematical results on linear as well as non-linear CA rules are derived. Integers associated with linear CA rules are defined as linear numbers and the properties of these linear numbers are studied.

## 1 Introduction

An elementary two state Cellular Automata (CA) rule describes how to determine a Boolean value output based on some logical calculation from Boolean inputs [4, 9]. Such rules play a basic role in questions of complexity theory as well as the design of circuits and chips for digital computers. The properties of CA rules play a critical role in Cryptography, Mathematics, Social Science, Engineering etc. Many methods exist to represent Cellular Automata rules [1, 2, 5, 7, 8, 10, 11 and 12]. Earlier a recursive algorithm has been designed which generates in-equivalent CA rules of any neighbourhood from the four in-equivalent CA rules of one-neighbourhood [7]. In this paper, some important theorems on CA rules are discussed in section 2. In section 3 the structures of linear CA rules are studied using the notion of binary trees and also generating formulas for linear CA rules are given. Integers associated with linear rules are named as linear numbers and the rest of integers are named as non-linear numbers. The properties of these linear and non-linear numbers are studied and given in section 4.

## 2 Some Theorems on Cellular Automata Rules

Let B=$\{0, 1\}$ be a set and let $B^n = B \times B \times \cdots \times B$ ($n$-times) is the Cartesian Product of B by $n$-times, then the set $B^n = \{(\alpha_1, \alpha_2, \ldots, \alpha_n) / \alpha_i \in B, \text{ for } 1 \leq i \leq n\}$. The $n$-neighbourhood CA rule denoted by $f_R^n$ (or $f$) is a function from $B^n$ to B defined as $f_R^n(\alpha_1, \alpha_2, \ldots, \alpha_n) = x_{2^{n-1}\alpha_1 + 2^{n-2}\alpha_2 + \ldots + \alpha_n} = x_k$, where $\alpha_i \in \{0,1\} \forall i = 1, 2, \ldots, n$ and $x_k \in \{0,1\}$ for $k = 2^{n-1}\alpha_1 + 2^{n-2}\alpha_2 + \ldots + \alpha_n$ and the rule number $R = (2^{2^n} - 1)x_0 + (2^{2^n} - 2)x_1 + \ldots + 2x_{n-1} + x_n$.

That is using the truth table representation of CA rule $f_R^n = \begin{pmatrix} x_0 \\ \ldots \\ x_{2^{2^n}-1} \end{pmatrix}$. The complement of a CA rule R is defined as $\overline{f}_R^n(\alpha_1, \alpha_2, \ldots, \alpha_n) = 1 \oplus x_k$, $0 \leq k \leq 2^n - 1$ and $\oplus$ is the binary EX-OR operator. CA rules are of two types linear and non-linear. A CA rule $f_R^n$ is said to be linear if it satisfies the linearity property that is $f_R^n(x \oplus y) = f_R^n(x) \oplus f_R^n(y)$ otherwise it is a non-linear CA rule. Linear CA rule in Algebraic Normal Form (ANF) [1, 4] is of the form $a_1 x_1 \oplus \ldots \ldots \oplus a_n x_n$ where $a_i$ is either 0 or 1 for $1 \leq i \leq n$. There are $2^{2^n}$ number of $n$-neighbourhood CA rules out of which the number of linear CA rules is $2^n$ and rest are all non-linear rules.

**Theorem 1:** $x_0 = 0$ is a necessary condition for $f$ to be a linear CA rule.
**Proof:** Given $f(0,0,\ldots,0) = x_0$. Linearity of $f$ implies that $f(0,0,\ldots,0) = f(0,0,\ldots,0) \oplus f(0,0,\ldots,0)$ that is
$x_0 = x_0 \oplus x_0$ ………………………………………………………………………………………………...........................(1)

We claim that $x_0 = 0$. If not then let $x_0 = 1$. Then from (1), we get $1 = 1 \oplus 1 = 0$. This contradicts the distinctness of 1 and 0. Hence $x_0 = 0$.

**Corollary 1:** If $x_0 = 1$ then CA rule is non-linear.
**Proof:** If not then let CA rule is linear. The linearity of CA rule implies that $x_0$ is necessarily zero, which contradicts the given condition that $x_0 = 1$.

**Definition:** Given an *n*-neighbourhood CA rule $f$ define two (*n+1*) neighbourhood CA rules $g$ and $h$ as follows
$g(0, \alpha_2, \alpha_3, ..., \alpha_{n+1}) = f(\alpha_2, \alpha_3, ..., \alpha_{n+1})$
$g(1, \alpha_2, \alpha_3, ..., \alpha_{n+1}) = f(\alpha_2, \alpha_3, ..., \alpha_{n+1})$
$h(0, \alpha_2, \alpha_3, ..., \alpha_{n+1}) = f(\alpha_2, \alpha_3, ..., \alpha_{n+1})$
$h(1, \alpha_2, \alpha_3, ..., \alpha_{n+1}) = \overline{f}(\alpha_2, \alpha_3, ..., \alpha_{n+1}) = 1 \oplus f(\alpha_2, \alpha_3, ..., \alpha_{n+1})$

**Representation of Rules $g$ and $h$:**
The CA rule $g$ as defined above is an (*n+1*)-neighbourhood CA rule which can be looked upon as the concatenation of the CA rule $f$ with itself, whereas $h$ is the concatenation of $f$ with its complement $\overline{f}$ with the constraint that $f$ shall always precede $\overline{f}$. Thus we will denote these two rules as $g = f f$ and $h = f \overline{f}$.

For example if $f = \begin{pmatrix} 0 \\ 0 \end{pmatrix}$, then $g = \begin{pmatrix} 0 \\ 0 \\ 0 \\ 0 \end{pmatrix}$ and $h = \begin{pmatrix} 0 \\ 0 \\ 1 \\ 1 \end{pmatrix}$ because $\overline{f} = \begin{pmatrix} 1 \\ 1 \end{pmatrix}$.

**Theorem 2:** $f$ is linear if and only if $g$ is linear.
**Proof:** To show the linearity of $g$ we need to show that
$g(\alpha_1 \oplus \beta_1, \alpha_2 \oplus \beta_2, \alpha_3 \oplus \beta_3, ..., \alpha_{n+1} \oplus \beta_{n+1}) = g(\alpha_1, \alpha_2, \alpha_3, ..., \alpha_{n+1}) \oplus g(\beta_1, \beta_2, \beta_3, ..., \beta_{n+1})$

**CASE 1:** $\alpha_1 = 0$ and $\beta_1 = 0$
$g(\alpha_1 \oplus \beta_1, \alpha_2 \oplus \beta_2, \alpha_3 \oplus \beta_3, ..., \alpha_{n+1} \oplus \beta_{n+1})$
$= g(0, \alpha_2 \oplus \beta_2, \alpha_3 \oplus \beta_3, ..., \alpha_{n+1} \oplus \beta_{n+1})$
$= f(\alpha_2 \oplus \beta_2, \alpha_3 \oplus \beta_3, ..., \alpha_{n+1} \oplus \beta_{n+1})$ (*defininition of g*)
$= f(\alpha_2, \alpha_3, ..., \alpha_{n+1}) \oplus f(\beta_2, \beta_3, ..., \beta_{n+1})$ (*as f is linear*)
$= g(0, \alpha_2, \alpha_3, ..., \alpha_{n+1}) \oplus g(0, \beta_2, \beta_3, ..., \beta_{n+1})$ (*defininition of g*)
$= g(\alpha_1, \alpha_2, \alpha_3, ..., \alpha_{n+1}) \oplus g(\beta_1, \beta_2, \beta_3, ..., \beta_{n+1})$

**CASE 2:** $\alpha_1 = 0$ and $\beta_1 = 1$ or $\alpha_1 = 1$ and $\beta_1 = 0$
Let $\alpha_1 = 0$ and $\beta_1 = 1$
$g(\alpha_1 \oplus \beta_1, \alpha_2 \oplus \beta_2, \alpha_3 \oplus \beta_3, ..., \alpha_{n+1} \oplus \beta_{n+1})$
$= g(0 \oplus 1, \alpha_2 \oplus \beta_2, \alpha_3 \oplus \beta_3, ..., \alpha_{n+1} \oplus \beta_{n+1})$
$= g(1, \alpha_2 \oplus \beta_2, \alpha_3 \oplus \beta_3, ..., \alpha_{n+1} \oplus \beta_{n+1})$
$= f(\alpha_2 \oplus \beta_2, \alpha_3 \oplus \beta_3, ..., \alpha_{n+1} \oplus \beta_{n+1})$ (*defininition of g*)
$= f(\alpha_2, \alpha_3, ..., \alpha_{n+1}) \oplus f(\beta_2, \beta_3, ..., \beta_{n+1})$ (*as f is linear*)
$= g(0, \alpha_2, \alpha_3, ..., \alpha_{n+1}) \oplus g(1, \beta_2, \beta_3, ..., \beta_{n+1})$ (*defininition of g*)
$= g(\alpha_1, \alpha_2, \alpha_3, ..., \alpha_{n+1}) \oplus g(\beta_1, \beta_2, \beta_3, ..., \beta_{n+1})$

Similarly it can be proved when $\alpha_1 = 1$ and $\beta_1 = 0$

**CASE 3:** $\alpha_1 = 1$ and $\beta_1 = 1$

$g(\alpha_1 \oplus \beta_1, \alpha_2 \oplus \beta_2, \alpha_3 \oplus \beta_3, ..., \alpha_{n+1} \oplus \beta_{n+1})$
$= g(1 \oplus 1, \alpha_2 \oplus \beta_2, \alpha_3 \oplus \beta_3, ..., \alpha_{n+1} \oplus \beta_{n+1})$
$= g(0, \alpha_2 \oplus \beta_2, \alpha_3 \oplus \beta_3, ..., \alpha_{n+1} \oplus \beta_{n+1})$
$= f(\alpha_2 \oplus \beta_2, \alpha_3 \oplus \beta_3, ..., \alpha_{n+1} \oplus \beta_{n+1})$     (*defininition of g*)
$= f(\alpha_2, \alpha_3, ..., \alpha_{n+1}) \oplus f(\beta_2, \beta_3, ..., \beta_{n+1})$     (*as f is linear*)
$= g(1, \alpha_2, \alpha_3, ..., \alpha_{n+1}) \oplus g(1, \beta_2, \beta_3, ..., \beta_{n+1})$     (*defininition of g*)
$= g(\alpha_1, \alpha_2, \alpha_3, ..., \alpha_{n+1}) \oplus g(\beta_1, \beta_2, \beta_3, ..., \beta_{n+1})$

Hence if $f$ is linear then $g$ is also linear.

Now given $g$ is linear then from the definition of $g$, it can be proved that $f$ is a linear CA rule that is to prove
$f(\alpha_2 \oplus \beta_2, \alpha_3 \oplus \beta_3, ..., \alpha_{n+1} \oplus \beta_{n+1}) = f(\alpha_2, \alpha_3, ..., \alpha_{n+1}) \oplus f(\beta_2, \beta_3, ..., \beta_{n+1})$
From (2) $f(\alpha_2 \oplus \beta_2, \alpha_3 \oplus \beta_3, ..., \alpha_{n+1} \oplus \beta_{n+1}) = g(0, \alpha_2 \oplus \beta_2, \alpha_3 \oplus \beta_3, ..., \alpha_{n+1} \oplus \beta_{n+1})$
$= g(0, \alpha_2, \alpha_3, ..., \alpha_{n+1}) \oplus g(0, \beta_2, \beta_3, ..., \beta_{n+1})$ (as $g$ is linear)
$= f(\alpha_2, \alpha_3, ..., \alpha_{n+1}) \oplus f(\beta_2, \beta_3, ..., \beta_{n+1})$. Hence proved.

**Theorem 3** $f$ is linear if and only if $h$ is linear.
**Proof:** To show the linearity of $h$ we need to show that
$$h(\alpha_1 \oplus \beta_1, \alpha_2 \oplus \beta_2, \alpha_3 \oplus \beta_3, ..., \alpha_{n+1} \oplus \beta_{n+1}) = h(\alpha_1, \alpha_2, \alpha_3, ..., \alpha_{n+1}) \oplus h(\beta_1, \beta_2, \beta_3, ..., \beta_{n+1})$$

**CASE 1:** $\alpha_1 = 0$ and $\beta_1 = 0$

$h(\alpha_1 \oplus \beta_1, \alpha_2 \oplus \beta_2, \alpha_3 \oplus \beta_3, ..., \alpha_{n+1} \oplus \beta_{n+1})$
$= h(0, \alpha_2 \oplus \beta_2, \alpha_3 \oplus \beta_3, ..., \alpha_{n+1} \oplus \beta_{n+1})$
$= f(\alpha_2 \oplus \beta_2, \alpha_3 \oplus \beta_3, ..., \alpha_{n+1} \oplus \beta_{n+1})$     (*defininition of h*)
$= f(\alpha_2, \alpha_3, ..., \alpha_{n+1}) \oplus f(\beta_2, \beta_3, ..., \beta_{n+1})$     (*as f is linear*)
$= h(0, \alpha_2, \alpha_3, ..., \alpha_{n+1}) \oplus h(0, \beta_2, \beta_3, ..., \beta_{n+1})$     (*defininition of h*)
$= h(\alpha_1, \alpha_2, \alpha_3, ..., \alpha_{n+1}) \oplus h(\beta_1, \beta_2, \beta_3, ..., \beta_{n+1})$

**CASE 2:** $\alpha_1 = 0$ and $\beta_1 = 1$ or $\alpha_1 = 1$ and $\beta_1 = 0$
Let $\alpha_1 = 0$ and $\beta_1 = 1$
$h(\alpha_1 \oplus \beta_1, \alpha_2 \oplus \beta_2, \alpha_3 \oplus \beta_3, ..., \alpha_{n+1} \oplus \beta_{n+1})$
$= h(0 \oplus 1, \alpha_2 \oplus \beta_2, \alpha_3 \oplus \beta_3, ..., \alpha_{n+1} \oplus \beta_{n+1})$
$= h(1, \alpha_2 \oplus \beta_2, \alpha_3 \oplus \beta_3, ..., \alpha_{n+1} \oplus \beta_{n+1})$
$= 1 \oplus f(\alpha_2 \oplus \beta_2, \alpha_3 \oplus \beta_3, ..., \alpha_{n+1} \oplus \beta_{n+1})$     (*defininition of h*)
$= 1 \oplus f(\alpha_2, \alpha_3, ..., \alpha_{n+1}) \oplus f(\beta_2, \beta_3, ..., \beta_{n+1})$     (*as f is linear*)
$= f(\alpha_2, \alpha_3, ..., \alpha_{n+1}) \oplus 1 \oplus f(\beta_2, \beta_3, ..., \beta_{n+1})$     (*commutative and associativity*)
$= h(0, \alpha_2, \alpha_3, ..., \alpha_{n+1}) \oplus h(1, \beta_2, \beta_3, ..., \beta_{n+1})$     (*defininition of h*)
$= h(\alpha_1, \alpha_2, \alpha_3, ..., \alpha_{n+1}) \oplus h(\beta_1, \beta_2, \beta_3, ..., \beta_{n+1})$

Similarly it can be proved when $\alpha_1 = 1$ and $\beta_1 = 0$

**CASE 3:** $\alpha_1 = 1$ and $\beta_1 = 1$

$h(\alpha_1 \oplus \beta_1, \alpha_2 \oplus \beta_2, \alpha_3 \oplus \beta_3, ..., \alpha_{n+1} \oplus \beta_{n+1})$
$= h(1 \oplus 1, \alpha_2 \oplus \beta_2, \alpha_3 \oplus \beta_3, ..., \alpha_{n+1} \oplus \beta_{n+1})$
$= h(0, \alpha_2 \oplus \beta_2, \alpha_3 \oplus \beta_3, ..., \alpha_{n+1} \oplus \beta_{n+1})$
$= f(\alpha_2 \oplus \beta_2, \alpha_3 \oplus \beta_3, ..., \alpha_{n+1} \oplus \beta_{n+1})$ \quad (*defininition of h*)
$= f(\alpha_2, \alpha_3, ..., \alpha_{n+1}) \oplus f(\beta_2, \beta_3, ..., \beta_{n+1})$ \quad (*as f is linear*)
$= 0 \oplus f(\alpha_2, \alpha_3, ..., \alpha_{n+1}) \oplus f(\beta_2, \beta_3, ..., \beta_{n+1})$
$= (1 \oplus 1) \oplus f(\alpha_2, \alpha_3, ..., \alpha_{n+1}) \oplus f(\beta_2, \beta_3, ..., \beta_{n+1})$
$= (1 \oplus f(\alpha_2, \alpha_3, ..., \alpha_{n+1})) \oplus (1 \oplus f(\beta_2, \beta_3, ..., \beta_{n+1}))$ \quad (*commutative and associativity*)
$= h(1, \alpha_2, \alpha_3, ..., \alpha_{n+1}) \oplus h(1, \beta_2, \beta_3, ..., \beta_{n+1})$ \quad (*defininition of h*)
$= h(\alpha_1, \alpha_2, \alpha_3, ..., \alpha_{n+1}) \oplus h(\beta_1, \beta_2, \beta_3, ..., \beta_{n+1})$

Hence if $f$ is linear then $h$ is also linear.

Now given $h$ is linear then from the definition of $h$,
$f(\alpha_2 \oplus \beta_2, \alpha_3 \oplus \beta_3, ..., \alpha_{n+1} \oplus \beta_{n+1}) = h(0, \alpha_2 \oplus \beta_2, \alpha_3 \oplus \beta_3, ..., \alpha_{n+1} \oplus \beta_{n+1})$
$= h(0, \alpha_2, \alpha_3, ..., \alpha_{n+1}) \oplus h(0, \beta_2, \beta_3, ..., \beta_{n+1})$ (as $h$ is linear)
$= f(\alpha_2, \alpha_3, ..., \alpha_{n+1}) \oplus f(\beta_2, \beta_3, ..., \beta_{n+1})$. Hence if $h$ is linear then $f$ is also linear.

From theorem 3 and 4 we conclude that
$$f \text{ is linear if and only if } ff \text{ and } f\overline{f} \text{ are linear}$$

From the above we have the following observations:

**Observation 1:** Linear CA rule concatenated with non-linear CA rule is again Linear CA rule provided the non–linear CA rule is the complement with which it is concatenated.

**Observation 2:** Linear CA rule concatenated with it self is a linear CA rule.

**Observation 3:** Rest of all concatenations gives new non-linear CA rule.

**Theorem 4:** The set of linear CA rules is not closed under concatenation operation.
**Proof:** As an example, the concatenation of two linear CA rules $[0\ 1]^T$ and $[0\ 0]^T$ is $[0\ 1\ 0\ 0]^T$ which is not linear.

## 3 Generation of linear CA rules

A recursive formula to generate n-neighborhood linear CA rule using truth table representation of CA rule is as $f_k^{n-1} \to [f_{(2^{2^{n-1}}+1)k}^n, f_{(2^{2^{n-1}}-1)(k+1)}^n]$ for $k = 0,1,2,...$ and $n = 1,2,3,...$ with initial function being $f_0^0 = 0$ for $n = 1$ and $k = 0$. The binary tree for this infinite recursion is shown in fig.1.

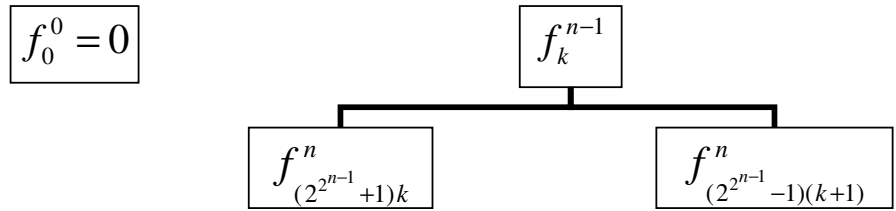

Fig 1: Shows the recursive binary tree for the recursive formula to generate linear CA rules.

As an example, fig.2 shows the complete binary tree of height 3.

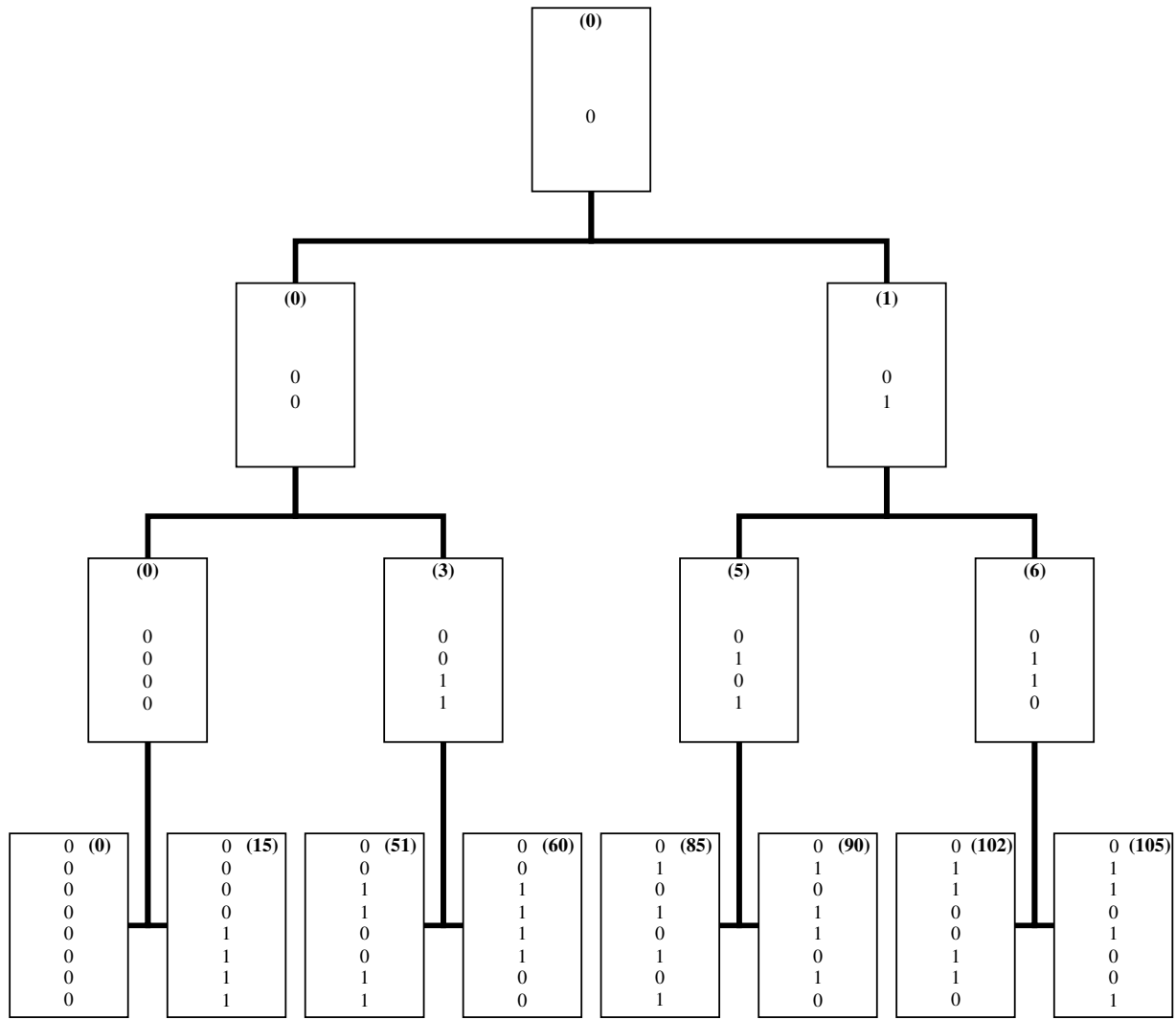

Fig 2: Shows the recursive binary tree that generates all linear CA rules upto 3-neighbourhoods, starting from a single node 0 and the linear rules are shown using truth table representation of CA rules.

The linear CA rules so generated uses truth table representation of CA rules but, when these rules are represented using Algebraic Normal Form (ANF) then the corresponding tree can be represented using the following recursive formula.

$$X_0^0 = 0,\ X_0^n = 0^{2^n}$$
$$X_{2^k}^{n-1} \rightarrow [X_{2^k}^n, X_{2^k}^n \oplus X_{2^{k+1}}^n]\ for\ k = 0,1,2,...and\ n = 1,2,3,...$$

Where $X_{2^k}^n$ is an $n$-neighbourhood CA rule of length $2^n$ and is a binary string of the form $0^{2^k}1^{2^k}...0^{2^k}1^{2^k}$.

Such representation helps in designing the Cellular Automata Machine [6 and 3].

## 4 Linear Numbers

**Definition:** Numbers associated with linear CA rules are called linear numbers and all other numbers are called non-linear i.e. if $f_R^n$ is a linear CA rule then $R$ is the linear number.

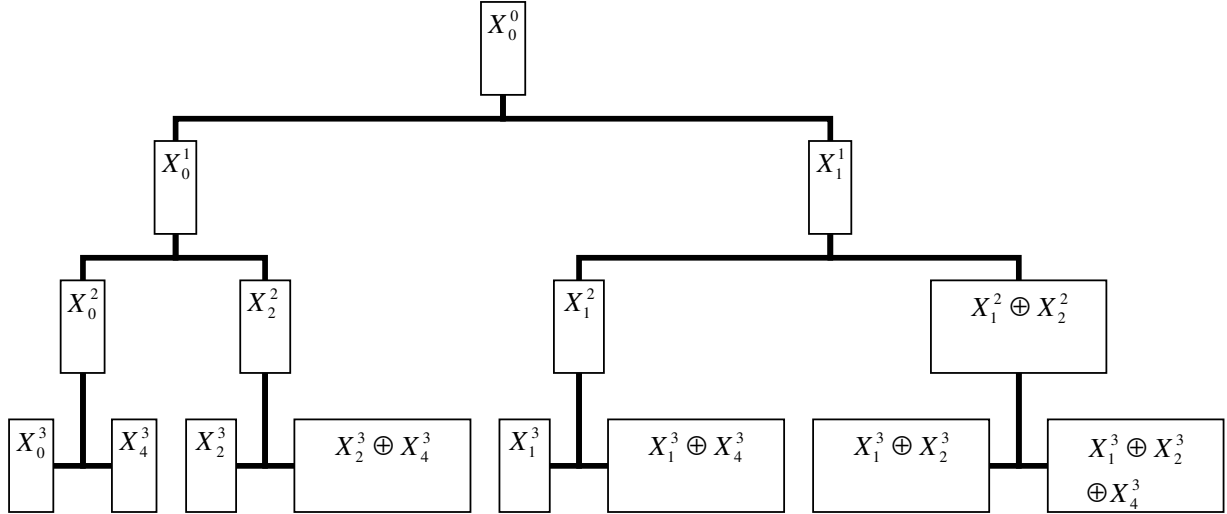

Fig 3: Shows the binary tree that generates all linear CA rules upto 3-neighbourhoods represented by their ANF.

## 4.1 Properties of linear numbers

1. From generating principle for linear CA rules if $R_n$ is a linear number then $(2^{2^{n-1}}+1)R_n$ and $(2^{2^{n-1}}-1)(R_n+1)$ are linear numbers.
2. Linear number corresponds to a CA rule appears in branches of binary trees. Thus on using the generating principle as shown in fig-1 and fig. 2, the set of linear numbers are 0, 3, 5, 6, 15, 51, 60, 85, 90, 102, 105 and so on.
3. At the height $n$, the number of such linear numbers is same as the number of linear CA rules which is $2^n$ and the length of each such number is thus $2^n$.
4. The binary string consisting of all 1's having length $2^k$ is a linear number at height $(k+1)$ and its value is $(2^{2^k}-1)$.
5. The set of all $n$-neighbourhood linear CA rules is a commutative group under the $\oplus$ operation.
6. First generation linear numbers, Second generation linear numbers etc. and accordingly they form an equivalence class with respect to its size at any level of the tree.

**Theorem 5: Every linear number will generate one even and one odd linear number.**
**Proof:** In the recursive formullae the terms $2^{2^{n-1}}+1$ and $2^{2^{n-1}}-1$ are necessarily odd numbers and let $k$ be a linear number. But, $k$ has two choices either it is even or it is odd. If k is odd then $(2^{2^{n-1}}+1)k$ is odd where as $(2^{2^{n-1}}-1)(k+1)$ is even. If k is even then $(2^{2^{n-1}}+1)k$ is even where as $(2^{2^{n-1}}-1)(k+1)$ is odd. Thus in both cases the successors of the linear number k produces one even and one odd linear number.

**Theorem 6: Linear numbers at any level of height $n$ in the recursive binary tree is always less than $2^{2^{n-1}}-1$**
**Proof:** If $k$ is a linear number at height $n$ then it is to prove that $k < 2^{2^{n-1}}-1$. Theorem-1 tells that $x_0 = 0$ is a necessary condition for a function to be a linear CA rule. Any binary string of length $2^n$ having MSB position is 0 is actually a binary string of length $(2^n - 1)$ and the largest decimal value of such binary string is $2^{2^{n-1}}-1$ that is when all $(2^n - 1)$ bit positions are 1, so all linear numbers must be less than this value.

**Theorem 7: The rule number associated with any CA rule x whose $x_0$ value 0 is always smaller than the number associated with its complement CA rule $x^c$.**

**Proof:** Let x be any k-bit CA rule whose $x_0$ value is 0. So the value of $x_0$ in its corresponding *k*-bit complement CA rule is 1. As $x_0$ is the most significant bit (MSB) position of both the k-bit string thus the string whose $x_0$ value is 0 is always smaller than the string whose $x_0$ value is 1. Hence proved.

**Corollary 2:** If x is a linear CA rule then the decimal value of x is less than the decimal value of $x^c$.
**Proof:** If x is linear then $x_0$ which is the MSB position of x is always 0. So by theorem 6 the decimal value of x is less than the decimal value of $x^c$.

**Corollary 3:** If x is a linear CA rule then decimal value of xx < decimal value of $xx^c$.
**Proof:** Easily follows from Theorem 7 and Corollary 2.

**Theorem 8:** The sequence of linear numbers are in strictly increasing order.
**Proof:** Easily follows from Corollary 3.

**Theorem 9: In the set of linear numbers the only prime numbers are 3 and 5.**
**Proof:** The numbers $(2^{2^{n-1}}+1)k$ and $(2^{2^{n-1}}-1)(k+1)$ are the product of two integers and hence they are composite except the numbers when one of the integer either k=1 in the first case or *k*+1=1 in the second case. This occurs only in two cases $(2^{2^{2-1}}-1)(0+1) = 3$ and $(2^{2^{3-1}}+1) \times 1 = 5$ which are primes.

# 5 Conclusion

In this paper, the detailed structure of linear CA rules is studied using a recursive binary tree. A generating formula for *n*-neighbourhood linear CA rule is given along with some of its properties. The set of integers are classified into linear and non-linear numbers based on the integers associated with linear and non-linear CA rules. The properties of these linear and non-linear numbers are studied. Questions that we are interested to solve in the future mainly includes: finding the distribution of all linear numbers in the set of all CA rules and finding how many linear numbers are there less than or equal to an integer x. The theory of linear and non-linear numbers may be useful in number theory as well as in non-linear dynamics.